\documentstyle[prc,aps,preprint,psfig]{revtex}
\begin{document}
\title{Off-shell effects on particle production\thanks{Work 
supported by DFG and GSI Darmstadt}}
\author{M. Effenberger and U.~Mosel\\
Institut f\"ur Theoretische Physik, Universit\"at Giessen\\
D-35392 Giessen, Germany\\
UGI-99-11}
\maketitle
\tightenlines
\begin{abstract}
We investigate the observable effects of off-shell propagation of nucleons in
heavy-ion collisions at SIS energies. 
Within a semi-classical BUU transport
model we find a strong enhancement of subthreshold particle production when
off-shell nucleons are propagated.    
\end{abstract}
\bigskip
During the past 15 years semi-classical transport models have extensively and
quite successfully been applied to the description of heavy-ion collisions
from SIS to SPS energies (see e.g. Refs.~\cite{brat,bass}). 
All these models are based on an on-shell approximation
for the nucleons whereas the nucleon resonances are propagated with their
spectral functions. However, already at SIS energies the collision rates are
so high that this approximation for the nucleons is hard to justify.
\par While the general formalism to treat also the off-shell transport of
particles has been available for quite some while 
\cite{kadbaym,daniel,botmal} only 
recently the interest in the in-medium properties of the 
$\rho$-meson has triggered some new efforts to develop a transport theory that
includes dynamical spectral functions (see e.g. \cite{knoll,cassing,knoll2}). 
In Ref.~\cite{photo} we have already presented a model in which the 
spectral functions
of the $\rho$ and $\omega$ mesons are dynamically incorporated and have 
discussed in detail the transport equation that we solve. 
It is the purpose of this letter to discuss the effects of collisional
broadening also on the spectral functions of the nucleons
and thus to treat them on the same footing as the resonances. We also want to
report first results on particle production in heavy-ion
collisions that were obtained by an inclusion of the nucleonic spectral 
function.
\par Since in our method a 'spectral phase space distribution' plays a
central role we specify its precise meaning by starting from  
the non-relativistic, semi-classical Kadanoff-Baym equation \cite{kadbaym}:
\begin{equation}
\left[p_0 - H_{mf},g^< \right]+
\left[{\rm Re}g,\Sigma^< \right] = -\Sigma^> g^< + \Sigma^< g^> \quad, 
\label{transport}
\end{equation} 
where the Poisson bracket is given as:
\begin{equation}
\left[ X, Y \right ]=\frac{\partial X}{\partial x_\mu}
\frac{\partial Y}{\partial p^\mu}-\frac{\partial X}{\partial p_\mu}
\frac{\partial Y}{\partial x^\mu} \quad,
\end{equation}
with $x$ and $p$ being the spatial and momentum coordinates. The mean field
Hamilton function $H_{mf}$ for the non-relativistic case reads:
\begin{equation}
\label{hamilton}
H_{mf}=\frac{\vec{p}^2}{2 M_0} + {\rm Re} \Sigma \quad,
\end{equation}
where $M_0$ denotes the mass of the particle and $\Sigma(x,p)$ its 
retarded self energy
that is related to the self energies $\Sigma^>$ and $\Sigma^<$ which appear 
in the 'collision term' on the rhs of Eq.~(\ref{transport}) via:
\begin{equation}
{\rm Im} \Sigma=\frac{1}{2} (\Sigma^> \mp \Sigma^<)
\end{equation}
with the upper sign for bosons and the lower sign for fermions, respectively.
\par $g^<(x,p)$ and $g^>(x,p)$ are the Wigner transforms of the real time 
Greens functions and $g^<$ can be interpreted as a generalized phase space 
density, i.e. the density of particles with 4-momentum $p$ at space time 
point $x$. The Wigner transform of the retarded Greens function $g(x,p)$ is
given by: 
\begin{equation}
{\rm Im} g = \frac{1}{2} (g^> \mp g^<) \equiv \frac{1}{2} a \quad,
\end{equation}
where $a(x,p)$ is the spectral function.
$g^<$ and $g^>$ can be expressed through the spectral function $a$ and
a phase space distribution function $f(x,p)$:
\begin{eqnarray}
g^< &=& af \\
g^> &=& a (1 \pm f).
\end{eqnarray}
For stable particles $f$ reduces to the usual phase space density.
From Eq.(\ref{transport}) and the corresponding equation for $g^>$ one obtains
an algebraic solution for $g$ \cite{kadbaym}:
\begin{equation}
g=\frac{1}{p_0-\frac{\vec{p}^2}{2 M_0} - \Sigma} .
\end{equation}   
The spectral function is then given as:
\begin{equation}
a=\frac{\Sigma^> \mp \Sigma^<}
{(p_0-\frac{\vec{p}^2}{2 M_0}-{\rm Re} \Sigma)^2+
\frac{1}{4} (\Sigma^> \mp \Sigma^<)^2}.
\label{spectral}
\end{equation}
For a thermally equilibrated system $f$ reduces to the usual Fermi-Dirac or
Bose-Einstein distribution function; the non-trivial spectral information
is then contained in $a$. 
\par In the following we will neglect the second Poisson bracket of 
${\rm Re}g$ and $\Sigma^<$ on 
the lhs of Eq.~(\ref{transport}). 
Its influence on the description of heavy-ion collisions is hard to
estimate because a numerical implementation of its effect is not available
so far. 
\par Instead of the non-relativistic Hamilton function Eq.~(\ref{hamilton}) we
use, in our specific numerical realization \cite{TeisZP97}, the following
one:
\begin{equation}
H_{mf}=\sqrt{(\mu+S)^2+\vec{p}^2} \quad,
\end{equation}
where $S(x,p)$ is an effective scalar potential and $\mu$ is a mass parameter 
that is related to the energy $p_0$ via:
\begin{equation}
p_0=H_{mf}.
\end{equation}
In the following we will express all quantities as function of 
$\mu$ instead of $p_0$.
We, therefore, instead of $g^<$ use a spectral distribution function 
$F(\vec{r},t,\vec{p},\mu)$ defined by:
\begin{equation}
F d\mu \equiv g^< \frac{d p_0}{2 \pi}
\end{equation}
and use instead of the spectral function from Eq.~(\ref{spectral}) the
following relativistic one:
\begin{equation} 
{\cal A}(\vec{r},t,\vec{p},\mu)=\frac{2}{\pi} 
\frac{\mu^2 \Gamma_{tot}(\vec{r},t,
\vec{p},\mu)}{(\mu^2-M_0^2)^2+\mu^2 \Gamma^2_{tot}(\vec{r},t,\vec{p},\mu)} 
\quad,
\label{spectral2}
\end{equation}          
where the total width in the rest frame of the particle $\Gamma_{tot}$ 
is given as:
\begin{equation}
\Gamma_{tot} = \frac{p_0}{\mu} (\Sigma^> \mp \Sigma^<).
\label{collwidth}
\end{equation}
The transport equation then reads:
\begin{equation}
(\frac{\partial}{\partial t} + \frac{\partial H_{mf}}{\partial \vec{p}} 
\frac{\partial}{\partial \vec{r}} - \frac{\partial H_{mf}}{\partial \vec{r}} 
\frac{\partial}{\partial \vec{p}})F=
\Sigma^< (1 \pm f) {\cal A} - \Sigma^> F.
\label{transport2}
\end{equation}
The collision term is given as:
\begin{eqnarray}
\lefteqn{\Sigma^< (1 \pm f) {\cal A} - \Sigma^> F = \frac{1}{2 E} \int 
\frac{d^3p_2 d\mu_2}{(2 \pi)^3 2 E_2} \
\frac{d^3p_3 d\mu_3}{(2 \pi)^3 2 E_3} \
\frac{d^3p_4 d\mu_4}{(2 \pi)^3 2 E_4} \
\times}\nonumber  \\
& &\times (2\pi)^4 \delta^{(4)}(p+p_2-p_3-p_4)  
\left| {\cal M} \right|^2  
\left\{ F_3 F_4 (1 \pm f_2){\cal A}_2 (1 \pm f) {\cal A} -\right.
\nonumber \\
& &  
\left.-F F_2 (1 \pm f_3) {\cal A}_3 (1 \pm f_4) {\cal A}_4 \right\} \quad,
\label{collterm}
\end{eqnarray}
where $F_i \equiv F(\vec{r},t,\vec{p}_i,\mu_i)$ and ${\cal A}_i$, $f_i$ are
denoted analogously.
\par We solve the transport equation~(\ref{transport2}) by a so-called
test particle method, i.e. the distribution function $F$ is discretized in
the following way:
\begin{equation}
F(\vec{r},t,\vec{p},\mu) \propto \sum_i \delta(\vec{r}-\vec{r}_i(t))
\delta(\vec{p}-\vec{p}_i(t)) \delta (\mu-\mu_i(t))  \quad,
\label{testpar}
\end{equation}
where $\vec{r}_i(t)$, $\vec{p}_i(t)$, and $\mu_i(t)$ denote the spatial
coordinate, the momentum and the mass of the test particle $i$, respectively.
By inserting this ansatz into the lhs of Eq.~(\ref{transport2}) one obtains
the usual equations of motion for the test particles between collisions:
\begin{eqnarray}
\frac{d \vec{r}_i}{d t}&=&
\frac{\partial H_{mf}}{\partial \vec{p}}
\\
\frac{d \vec{p}_i}{d t}&=&
-\frac{\partial H_{mf}}{\partial \vec{r}}
\label{eomp}
\\
\frac{d \mu_i}{d t}&=&0.
\label{eommu}
\end{eqnarray} 
\par A formal generalization of the transport equation to a system with
different particle species is straightforward. One obtains a transport
equation for each particle species that is coupled to all others via
the mean field potential and the collision term.
In our transport model all baryonic resonances up to about 2 GeV as well
as all low lying meson states are explicitly propagated. For a detailed
description of the model we refer to Refs.~\cite{photo,TeisZP97}. 
The actual collision term for the nucleons is thus much more general 
than the one
in Eq.~(\ref{collterm}) and a numerical solution is only possible by making
some approximations that we will describe in the following.
\par We neglect any medium modifications of total cross sections and
decay widths except of Pauli blocking of the outgoing nucleons. 
The cross sections are calculated as function of the 'free' invariant energy
$\sqrt{s}_{free}$ which is given as:
\begin{equation}
\sqrt{s}_{free}=\sqrt{\mu_1^2+p_{cm}^2}+\sqrt{\mu_2^2+p_{cm}^2} \quad,
\end{equation}
where $\mu_1$ and $\mu_2$ are the masses of the incoming particles and
$p_{cm}$ denotes their cms 3-momentum. For invariant energies below $2 M_N$
we assume the cross section to be constant. We have checked that our results
do not change when we use the elastic nucleon-nucleon cross section as
function of $\sqrt{s}_{free}-\mu_1-\mu_2$ instead of as function of 
$\sqrt{s}_{free}$.
The in-medium
spectral functions of the produced particles enter only in the determination
of the 4-momenta of the outgoing particles. For 2 particles in the final
state the masses and the production angle are chosen according to
\begin{equation}
\frac{dn}{d\Omega d\mu_3 d\mu_4} \propto W(\theta) p_f {\cal A}_3 {\cal A}_4
\quad,
\label{massdist}
\end{equation}
where $W(\theta)$ denotes the respective angular distribution 
and $p_f$ is the cms final
state momentum of the outgoing particles with masses $\mu_3$ and $\mu_4$. 
\par
\begin{figure}[h]
\centerline{
{\psfig{figure=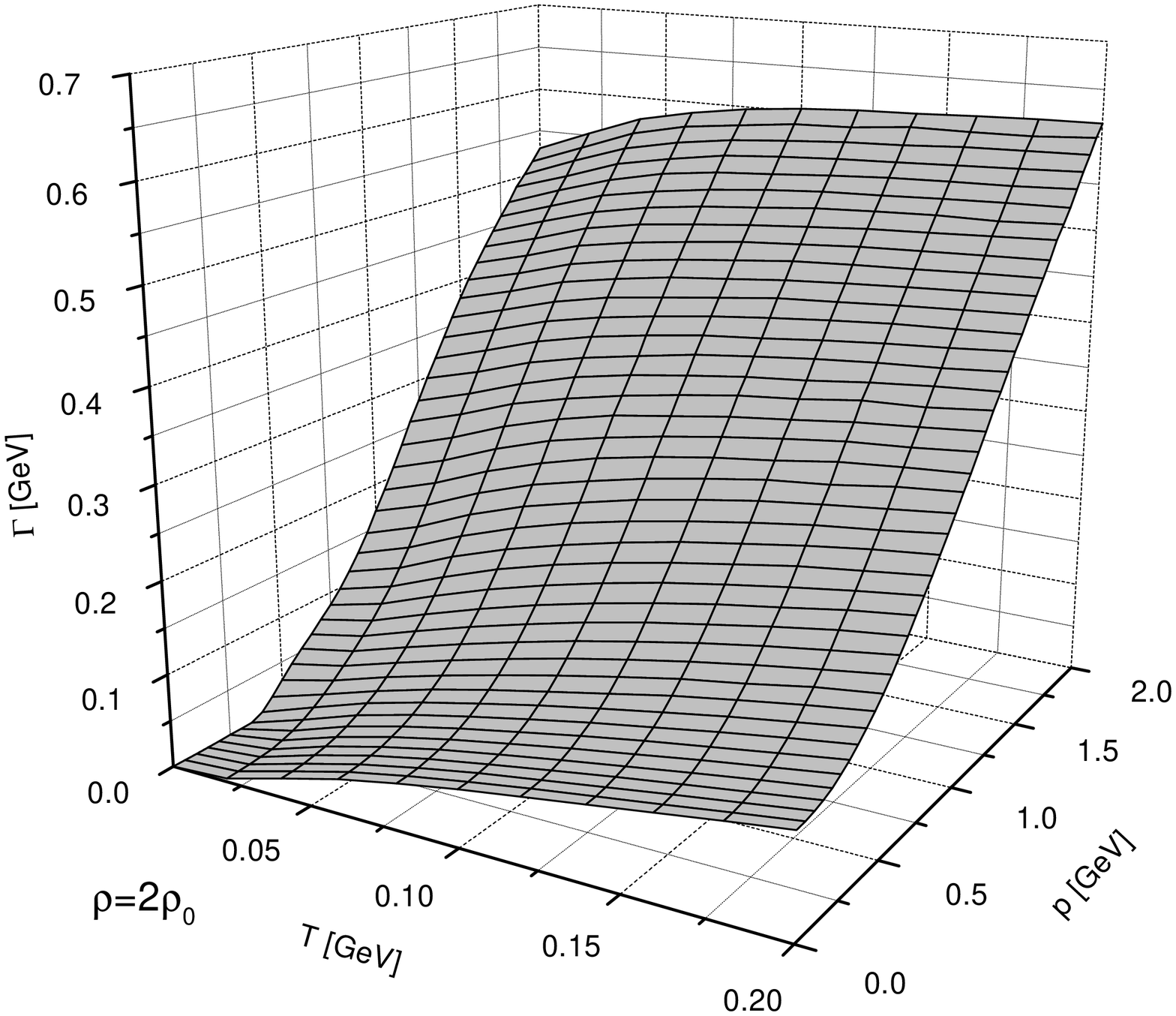,width=10cm}}}
\caption{The nucleon width as function of momentum $p$ and temperature $T$ at
density $\rho=2\rho_0$.}
\label{fig1}
\end{figure}

 Because in an actual simulation it is practically impossible to
obtain statistics for a determination of the collision rate in
8-dimensional phase space 
we calculate the collision width of the nucleons that enters 
the spectral functions in Eq.~(\ref{massdist}) only as function of 
total momentum in the
local rest frame $p_{lrf}$, density $\rho$ and temperature $T$:
\[
\Gamma_N(\vec{r},t,\vec{p},\mu) \to \Gamma_N(p_{lrf},\rho(\vec{r}(t)),
T(\vec{r}(t))).
\]
The temperature is determined by assuming a Fermi-Dirac distribution from
the local $<p^2>$ of the test particles. The width is calculated with
the nucleon-nucleon cross sections described in 
Ref.~\cite{TeisZP97} where we take into account Pauli
blocking for the elastic part (see e.g. Ref.~\cite{photo} for details on
the calculation of the collisional widths). In Fig.~\ref{fig1} we show the
width of the nucleons as function of momentum and temperature at density
$2\rho_0$. One sees that the width is of the order of 300 MeV for a 
temperature of 100 MeV and a momentum of 1 GeV and therefore
non-negligible. We have checked that the actual collision rates 
(time integrated and averaged over momentum)
in the
simulation agree very well with these widths. This holds also for the
early, non-equilibrium stages of the collision because the
collision rates depend mainly on the density and the momentum. 
We also find that it is essential to keep the rather strong momentum
dependence of the widths, depicted in Fig.~\ref{fig1}.    
\par In Ref.~\cite{photo} we have discussed in detail that the transport
equation Eq.~(\ref{transport2}) does not give the correct asymptotic solutions
for particles that are stable in vacuum  because a collision broadened 
particle does not automatically lose its collisional width when being 
propagated out of the nuclear environment.
In Ref.~\cite{photo} we have introduced a simple scheme to restore the
asymptotic width by adding a scalar potential to the equations
of motion for the test particles that shifts a particle on its mass shell
when it propagates to the vacuum. 
In order to fulfill energy conservation in case of a system of off-shell
nucleons we have to take into account the back coupling of the potential
on the other nucleons that create the potential via the density.
For a testparticle
$i$ this restoring potential is defined as:
\begin{equation}
s_i(\vec{x}_i(t),t))=(\mu_i(t_{cr})-\mu_{vac}-\bar{s}
(\vec{x}_i(t_{cr}),t_{cr}))
\frac{\rho(\vec{x}_i(t),t)}{\rho(\vec{x}_i(t_{cr}),t_{cr})} \quad,
\label{fakepot}
\end{equation} 
where $t_{cr}$ denotes the time when the particle is produced; for
nucleons we have
$\mu_{vac}=M_N$. The mass $\mu_i$
is then given as:
\begin{equation}
\mu_i(t)=\mu_{vac}+s_i(\vec{x}_i(t),t)+\bar{s}(\vec{x}_i(t),t).
\end{equation}
In addition to the scheme in Ref.~\cite{photo} we have included the back 
coupling potential $\bar{s}$ which is given as:
\begin{equation}
\bar{s}(\vec{r},t)=\frac{1}{2}\frac{
\int d^3 p d\mu (\mu-\mu_{vac}) F(\vec{r},t,\vec{p},\mu)}
{\int d^3 p d\mu F(\vec{r},t,\vec{p},\mu)}.
\label{backcoup}
\end{equation}
This prescription can be formulated in terms of a modification of the
equations of motion for the testparticles:
\begin{equation}
\frac{d \mu_i}{d t}=\frac{\mu_i-\mu_{vac}-\bar{s}}{\rho} 
\frac{d \rho}{d t} \approx \frac{\mu_i-\mu_{vac}-\bar{s}}{\Gamma_{coll}} 
\frac{d \Gamma_{coll}}{d t} \quad, 
\label{eommu2}
\end{equation} 
where in the last step we have assumed that the collisional width 
$\Gamma_{coll}(t) \propto \rho(t)$. 
Since $s_i$ and $\bar{s}$ enter the equations of
motion as usual potentials we also get a respective modification of 
Eq.~(\ref{eomp}):
\begin{equation}
\frac{d \vec{p}_i}{d t}=
-\frac{\mu+S}{p_0} \left( \frac{\partial S}{\partial \vec{r}}
+\frac{\mu-\mu_{vac}-\bar{s}}{\Gamma_{coll}}
\frac{\partial \Gamma_{coll}}{\partial \vec{r}} +
\frac{\partial \bar{s}}{\partial \vec{r}} \right).
\label{eomp2}
\end{equation} 
This procedure conserves total
energy since $\bar{s}$ is the local average of the $s_i$'s. 
Numerically we obtain energy conservation on a level better than 1\%
whereas
a neglect of the back coupling potential $\bar{s}$ leads to a violation
of total energy conservation by about 10\%. 
However, for the calculations of particle
yields presented here the inclusion of the restoring potential 
Eq.~(\ref{fakepot}) is
only of minor importance as will be shown below. 
Because of this it is not essential to what extent our prescription 
mimicks the effect of the Poisson bracket that we neglected in 
Eq.~(\ref{transport}). We note that the transport equation (\ref{transport})
is based on a first order gradient expansion and therefore
does not necessarily need to give the correct asymptotic solutions if the
gradients are too large. If the gradients are low enough already the
collision term shifts the particles on mass shell as we discussed in
Ref.~\cite{photo}. 
\par In the following we  
exemplarily present results of the present formalism
for central Au+Au collisions at 1 AGeV bombarding
energy. We have checked that the observed effects are similar for larger
impact parameters and smaller systems.
The results that we show here were obtained 
with an EOS that has a compressibility $K$ of 380 MeV. With a
softer EOS ($K$=215 MeV) the effects of off-shell nucleons are quantitatively
the same. In these calculations we employ a (somewhat arbitrary) lower
mass cut-off of 400 MeV to avoid unphysical test particle velocities. The
results are stable when this cutoff is varied.
\par
\begin{figure}[h]
\centerline{
{\psfig{figure=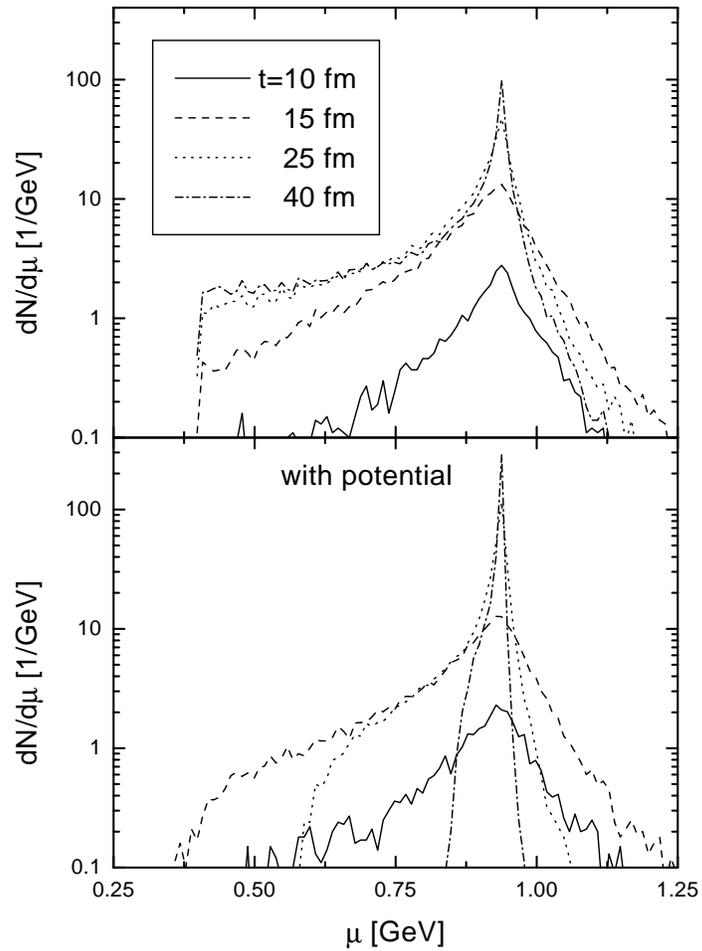,width=10cm}}}
\caption{Spectral distribution of nucleons for Au+Au at 1 AGeV. In the lower
part the restoring potential from Eq.~(\ref{fakepot}) was included (without the
backcoupling potential $\bar{s}$).}
\label{fig2}
\end{figure}

 In Fig.~\ref{fig2} we show the mass differential spectrum
of the nucleons for different times of the reaction:
\begin{equation}
\frac{dN}{d\mu}(t)= \int \frac{d^3 p}{(2 \pi)^3} d^3r 
F(\vec{r},t,\vec{p},\mu) \quad,  
\end{equation}   
where we only count the nucleons that suffered at least from one collision
since the others contribute to a $\delta$-function at the pole mass. The
upper part of the figure displays the result without the potential 
Eq.~(\ref{fakepot}). One sees that the nucleon distribution at the end
of the collision ($t=40$ fm) is quite strongly peaked around the nucleon pole
mass (FWHM$\approx$10 MeV). However, there is still some quite broad tail
of low mass nucleons. The inclusion of the potential Eq.~(\ref{fakepot}) 
'cures' this problem (lower part of Fig.~\ref{fig2}). One should note here
that the spectral distribution in the high density phase of the reaction is
hardly affected by the potential. The distribution has a width in the order
of 200 MeV which again stresses that an on-shell approximation for the
nucleons is hard to justify. It is an essential property of these spectral
distributions that they are strongly asymmetric, with an increase towards lower
masses. This stems mainly from available phase space which gets larger for
smaller masses even without the factor $p_f$ in Eq.~(\ref{massdist}). We 
stress again that this is a consequence of the collisions and not of the
potential~(\ref{fakepot}).
\par
\begin{figure}[h]
\centerline{
{\psfig{figure=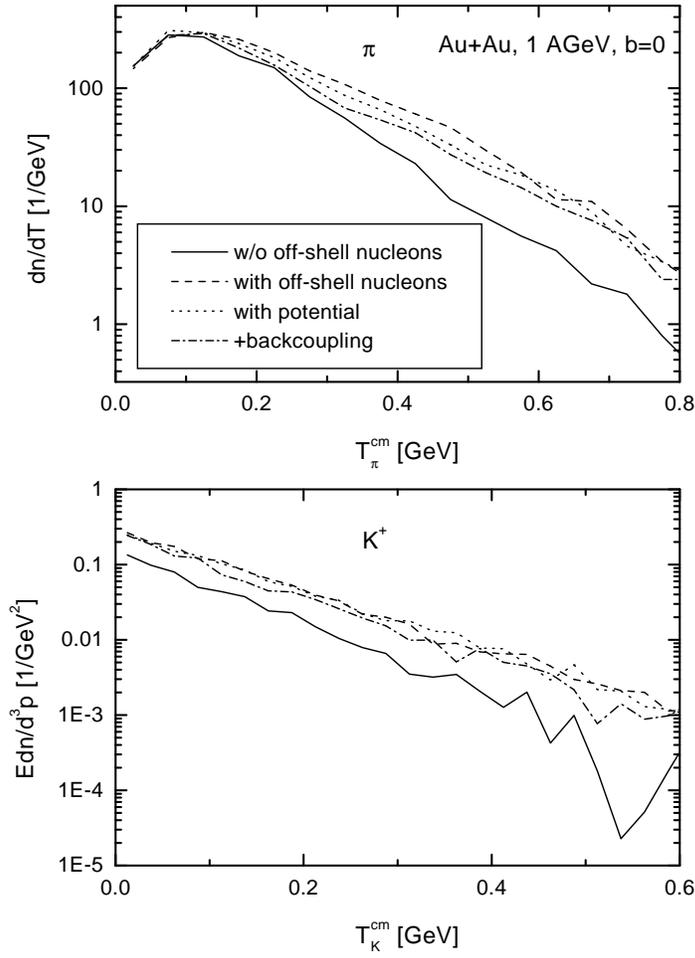,width=10cm}}}
\caption{Effect of off-shell propagation of nucleons on pion (upper part) and
kaon (lower part) spectra. For the dotted lines the restoring potential from 
Eq.~(\ref{fakepot}) was included. For the dash-dotted lines the 
energy-conserving backcoupling
potential Eq.~(\ref{backcoup}) was additionally taken into account.}
\label{fig3}
\end{figure}

 In Fig.~\ref{fig3} (upper part) we show that the off-shell
propagation has a quite large influence on the pion spectrum. 
The number of
high energy pions is increased by about a factor 2. There are essentially two
reasons for this result. First, in the initial stages of the collision high
mass nucleons can be produced which contribute to high energy secondary 
collisions. This is the effect which has been considered in 
Ref.~\cite{bertsch}. Secondly, outgoing nucleons can
have low masses so that effectively the 'temperature' increases. This
allows to produce heavier baryonic resonances in
$N N \to N R$ and more energetic pions in $R \to N \pi$.
We note that also in the quantummechanical off-shell investigation of 
Ref.~\cite{bozek} a higher particle yield is obtained when in-medium
correlations are taken into account.     
\par In the calculation of these results we have used the free 
$\Delta$ spectral
function; in the population the correct phase space factor has been taken
into account. We have ascertained that using the in-medium widths of
Ref.~\cite{abspaper} in the population of the $\Delta$ made no significant 
difference. 
\par The lower part of Fig.~\ref{fig3} shows our result for the 
$K^+$-spectrum where we see that the off-shell propagation gives an
enhancement by about a factor 2. This enhancement is solely caused by 
the increase of high energy baryon-baryon and pion-baryon collisions 
since we do not 
modify the total production cross sections. This would even lead to a
further increase of kaon production because in $B B \to N Y K$ a finite
width of the outgoing nucleon lowers effectively the production threshold.
\par The potential Eq.~(\ref{fakepot}) (with $\bar{s}$=0) 
has only a small effect both on the 
pion and on the kaon spectrum as can be seen by comparing the dashed and the
dotted curves in Fig.~\ref{fig3}. 
This is so because in the high density phase,
when the high energy mesons are produced, the collision rates for the
nucleons are so large that the nucleons do not travel through a relevant
density gradient for which the potential would become important. The potential
mainly shifts the low mass tail of nucleons on mass shell during later
stages of the collision. For subthreshold particle production the exact
mechanism how particles in heavy collision come to their mass shell plays
only a minor role. 
The calculation which
includes the backcoupling potential Eq.~(\ref{backcoup}) and thus conserves
total energy gives practically the same result (dash-dotted curves in
Fig.~\ref{fig3}).      
\par We shortly want to mention a few possible improvements for future work.
By using vacuum cross sections and vacuum decay widths, as it is always done in
transport calculations, detailed balance is violated when off-shell nucleons
are propagated. This could only be cured by working with amplitudes
instead of parameterizations of cross sections and decay widths which would
enhance the numerical effort dramatically. 
However, we do not expect the violation of detailed balance to be 
significant for 
our present results because the
harder particles for which the increase due to off-shell effects is most
pronounced (Fig.~\ref{fig3}) are produced in the first hard collisions.
Furthermore, in all transport calculations the analyticity of the 
self energy $\Sigma$ 
is neglected and the mean field potential that enters into the drift term 
is determined independently of
the self energies $\Sigma^<$ and $\Sigma^>$ in the collision term. For
off-shell transport this analyticity might be important for the mechanism
how the particles come to their mass shell when they travel to the vacuum.  
In principle, we should also take into account a dynamical spectral
function for the pion since its collision rates are even larger than the
ones of the nucleons.
\par In summary, we have presented a first calculation 
of nucleonic spectral functions in a transport theoretical framework and
have exploited their effects on particle
production in heavy-ion collisions at SIS energies. We have 
found large effects compared to a calculation with an on-shell approximation
for the nucleons. We thus hope that this work will stimulate future, more
refined work on the influence of off-shell properties on particle production
yields in heavy-ion collisions.
\bigskip      
\par The authors would like to thank C.~Greiner, J.~Knoll, V.~Koch, 
A.~La\-rio\-nov, and S.~Leu\-pold for 
discussions and helpful comments on the manuscript. They also acknowledge
a critical reading of the manuscript and lively arguments with W.~Cassing.

\
\newpage

\end{document}